\documentclass[11pt]{article}
\usepackage{jheppub}
\usepackage[usenames,dvipsnames,table]{xcolor}
\usepackage{graphicx,amsmath,amssymb,pgfplots,tikz}
\pgfplotsset{compat=1.18}
\newcommand{\F}[1]{\ignorespaces}
\title{Zassenhaus decomposition of half-sided translations and generalizations in 2d conformal field theory}
\author{Manish Ramchander}
\affiliation{The Institute of Mathematical Sciences,
IV Cross Road, C.I.T. Campus, Taramani, Chennai, India 600113}
\affiliation{Homi Bhabha National Institute, Training School Complex,
Anushakti Nagar, Mumbai, India 400094}
\emailAdd{manishd@imsc.res.in}

\definecolor{rust}{rgb}{0.9,0.2,0.2}

\F{}
\abstract{

    We study the half-sided translations associated to Rindler wedge algebras for conformal field theories in 1+1 Minkowski spacetime, generated by an unbounded operator $\mathcal{G}$, in terms of bilinear forms $G, G'$ made from entanglement Hamiltonians of the underlying algebras such that $\mathcal{G} = G+G'$. We show that despite entanglement Hamiltonians being ill-defined operators on Hilbert space, $G, G'$ can be regularized using smooth bump functions to operators  $\hat{G}, \hat{G}'$ with well-defined commutators, and use them to do a centered Zassenhaus expansion of $\exp(i \mathcal{G} s)$  in terms of $\hat{G}$ and $\hat{G}'$ which is tractable and respects causality. 
    We show that in fact half-sided translations is a special case in a large class of operators $\mathcal{O}$ for which a similar decomposition can be done by defining $\mathcal{O} = O_L+O_R$ with $O_{L}, O_{R}$ chosen approriately.

}

\begin{document}

\maketitle

\label{sec:intro}
\section{Introduction}

In the algebraic quantum field theory (AQFT), an important group of unitary transformations that can translate fields through the spacetime is given by the half-sided translations \cite{Borchers:1992cpt, Borchers:1998tvn}. As suggested by their name, these transformations are unlike the ordinary spacetime translations, for their definition requires specification of {\it two} von Neumann algebras $\mathcal{M}$ and $\mathcal{N}$ satisfying $\mathcal{N} \subset \mathcal{M}$, which together characterize them. It is with respect to these algebras, that the word ``translations" is used to describe the group, for when they exist, half of their group elements preserve the algebra $\mathcal{M}$, while the other half do not. Introduced in 1992 by Borchers in \cite{Borchers:1992cpt}, and studied at length thereafter,  the half-sided translations have enabled important advancements in AQFT \cite{Borchers:1993mis,  Wiesbrock:1993hsm, Borchers:1996hsi, Borchers:1998tvn, Borchers:1999qts, Borchers:2000pv}, and in the recent years, following the work of Leutheusser and Liu \cite{Leutheusser:2021frk, Leutheusser:2021qhd}, they have served as powerful tools in AdS/CFT for studying quantum field theories and quantum gravity \cite{Jalan:2023dmq, Krishnan:2023sii, Jalan:2024cby, Chen:2310dcw, Witten:2021gcp, Faulkner:2405gag}.

To what do the half-sided translations owe such a success in helping us understand quantum field theories, and how does one construct them, are two questions with the same answer -- the properties of modular Hamiltonians. The generator $\mathcal{G}$ of half-sided translations for the given pair $\mathcal{M}$ and  $\mathcal{N}$ is a function of their automorphism generators\footnote{A brief review is in section 2.}, known as the modular Hamiltonians, commonly denoted $K_{\mathcal{M}}, K_{\mathcal{N}}$ respectively \cite{Haag:1996hvx}.  To wit,
\begin{align}
\label{mathc}
\mathcal{G} = \frac{1}{2\pi} (K_{\mathcal{M}}- K_{\mathcal{N}}).
\end{align}
\noindent In the past decade, the role of modular Hamiltonians has been studied at length in AdS/CFT and in particular they have been shown to be of primary importance for bulk reconstruction \cite{Faulkner:2013lm, Faulkner:2017blmf, Jafferis:2016jlms,  Callebaut:2018xfu, Jafferis:2020ihr}. However, there is a qualification needed in this description that we must provide immediately, which will also allow us to introduce another key object for this work.

The high energy community often refers to the modular Hamiltonian as the ``full modular Hamiltonian", in contrast to the quantities called ``half-modular Hamiltonians" or often simply modular Hamiltonians, which are obtained by the logarithm of appropriate density operators. It is the latter, which have received most of the attention, and in condensed matter community these operators have been studied for many decades by the name entanglement Hamiltonians \cite{Dalmonte:2022eh, Cardy:2016fqc}. While this observation may seem frivolous, the point is that there is an indisputable difference between the notions of modular Hamiltonians and entanglement Hamiltonians, which is the following. $K_{\mathcal{M}}$, the modular Hamiltonian for the algebra $\mathcal{M}$, generates the automorphisms for $\mathcal{M}$ and its commutant $\mathcal{M}'$ both. In contrast, writing $K_{\mathcal{M}}  = H_{\mathcal{M}}- H_{\mathcal{M}'}$ in terms of entanglement Hamiltonians, the operators $H_{\mathcal{M}}$ and $H_{\mathcal{M}'}$ generate automorphisms of $\mathcal{M}$ and $\mathcal{M}'$ separately.  When focusing on properties such as the subregion duality, the natural object of study is the entanglement Hamiltonian, which is the reason why it has been preferred in problems concerning subregions of spacetime. The point is also that the half-sided translations, modular Hamiltonians, and the entanglement Hamiltonians are intimately connected, and in fact, in terms of the entanglement Hamiltonians,  one can split $\mathcal{G}$ as
\begin{align}
\label{split}
\mathcal{G}= G+G'; \quad G = \frac{1}{2\pi}( H_\mathcal{M} - H_{\mathcal{N}}),\quad G'= \frac{1}{2\pi}( H_{\mathcal{N}'} - H_{\mathcal{M}'}).
\end{align}

Having introduced the key objects of interest, let us now come to the premise of this work. The half-sided translations are able to translate a local operator $\phi(x,t)$ through unbounded regions in spacetime. This is because the involved modular Hamiltonians $K_{\mathcal{M}}, K_{\mathcal{N}}$ have a non-trivial effect on elements of both the algebras $\mathcal{M}, \mathcal{N}$ and their commutants. In contrast, the action of an entanglement Hamiltonian $H_{\mathcal{M}}$ on a local operator $\phi$ placed in the causal diamond $\Diamond_{\mathcal{M}}$ associated to $\mathcal{M}$ cannot take it outside $\Diamond_{\mathcal{M}}$, simply because $H_{\mathcal{M}}$ generates the automorphism of $\mathcal{M}$. The operators $G, G'$ similarly cannot be expected to transport  $\phi(x,t)$ through unbounded regions in spacetime either. One may then ask, how is it that $G$ and $G'$ combine together to create the effect of $\mathcal{G}$. We can make this language more precise as follows. Since  $\mathcal{G}$ is the generator of half-sided translations, the group elements are given by  $\exp (i \mathcal{G} s)$ for $s \in \mathbb{R}$. One may ask whether it is possible to understand this group element by ``stitching" the effect of  $e^{i G s}$ and $e^{iG's}$ in some way.

The natural intuition from experience in quantum mechanics is to consider the Zassenhaus expansion of $e^{i \mathcal{G} s}$ \cite{Magnus:1954ex, Kimura:2017edz}, using the nested commutators of $G$ and $G'$. One may therefore hope that an expansion of the form
\begin{align}
\label{hope}
e^{is \mathcal{G}} = e^{is G'} e^{is G} \cdot \prod_{k=2}^{\infty}  e^{(is)^{k}C_k} \quad (?),
\end{align}
\noindent with every $C_k$ made from nested commutators of  $G,G'$, will provide the answer. However, even if one ignores the difficulty of computing the infinite set $\{C_k\}$ for the moment, there is an issue that $\mathcal{G}$ is generally an unbounded operator.  Usually Zassenhaus decompositions involve exponentials of bounded operators on both sides, which can always be defined on the full Hilbert space.  Fortunately $\mathcal{G}$ is self-adjoint, thus $e^{i s \mathcal{G}}$ is unitary and therefore defined on the full Hilbert space $\mathcal{H}$. However, the status of the right side of \eqref{hope} is questionable, in whether the decomposition works on any subspace of the Hilbert space. \footnote{This expansion can indeed fail; cf. \ref{sec:zas}.}

Importantly, the suggestion \eqref{hope} faces a major objection which is the fact that entanglement Hamiltonians $\{H_{\mathcal{M}}, H_{\mathcal{N}}, H_{\mathcal{M}'}, H_{\mathcal{N}'}\}$ and therefore $G,G'$ along with $\{C_k\}$ are ill-defined operators because of the continuum nature of quantum field theory \cite{Witten:2018zxz}. This phenomenon is related to the fact that entanglement Hamiltonians cannot be regarded as logarithms of density matrices defined on separate Hilbert spaces in a strict sense because type III$_1$ algebras do not admit a trace \cite{Sorce:2023typ}.
An ill-defined operator here means that one whose matrix elements for suitable states may be well-defined, but as a map between states with a unique adjoint, it cannot be defined on any dense set,\footnote{Dense definition is required for a unique adjoint \cite{Reed:1981fa}. Put precisely, they are not operators, but are studied using unbounded bilinear forms, or operator valued distributions, and quantum fields fall in the same class. See section \ref{sec:g-g'} for more details.} that is, the mapped vector always has infinite norm. Consequently, even if one can compute the commutators of $G$ and $G'$,
%using stress tensor commutators -- which can be done for 2d CFTs using an $i\epsilon$ prescription of \cite{Besken:2021oli} --
the right side of \eqref{hope} will be an ill-defined operator which can be meaningful in its matrix elements for states in some dense subset of the Hilbert space at its best.

In this work, we will show this difficulty is surmountable by a regularization procedure. We will see that is possible to trade $G, G'$ for operators $\hat{G}, \hat{G}'$ obeying $\hat{G} + \hat{G}' = \mathcal{G}$ and mimicking the former, such that all the nested commutators of $\hat{G}, \hat{G}'$ are well-defined operators. Consequently we will show that $e^{i \mathcal{G} s}$ can be decomposed using $\hat{G}, \hat{G}'$, not in a right-sided, but a centered Zassenhaus expansion of the form:

\begin{align}
\label{centr}
e^{is \mathcal{G}} = e^{is \hat{{G}}'} \cdot  \prod_{k=2}^{\infty}    e^{(is)^{k}D_k} \cdot  e^{is \hat{G}}
\end{align}%
where $\{D_k\}$ are made from nested commutators of $\hat{G}, \hat{G}'$, with $D_k = \lambda_k D(\epsilon) + \Sigma_k(\epsilon)$; $\lambda_k$ is a $c$-number, $\epsilon$ is a regularization parameter, $\Sigma_k(\epsilon)$ are contributions arising from the associated regularization, and importantly, $D(\epsilon)$, the primary component of $D_k$,  is independent of $k$. We will in fact show that half-sided translations can be a special case in a large class of operations that permit such a Zassenhaus decomposition.

More specifically, we will be working with a general 2d conformal field theory in Minkowski spacetime, taking half-sided translations for $\mathcal{M}$ corresponding to the Rindler wedge as our prototypical example, and obtain the following results:

\begin{itemize}
\item We will show that norms of states ${H}_{\mathcal{M}}|\chi\rangle$ for generic CFT states $|\chi\rangle$ have a universal ultraviolet divergence coming from the boundary of the associated integral as suggested in \cite{Witten:2018zxz}, but also a universal infrared divergence. Contrarily, the states $G|\chi \rangle$ will only have a universal ultraviolet divergence that is marginal, and is easy to regularize. This will allow us to trade $G, G'$ for $\hat{G}, \hat{G}'$ by a boundary modification of the integral, such that  $\hat{G}|\chi \rangle, \hat{G}'|\chi\rangle$ have a finite norm. (Section \ref{sec:g-g'}).

\item The operators $\exp( is \hat{G} )$ and $\exp( is \hat{G}' )$ will be shown to be equivalent to $\exp( i s \mathcal{G} )$ in regions of spacetime causally disconnected to  $\hat{G}'$ and $\hat{G}$ respectively. However  $\exp( i s \hat{G} )$, $\exp( i s \hat{G}' )$ will turn out to be non-unitary, and therefore $\hat{G}, \hat{G}'$ will not be self-adjoint. (Section \ref{sec:conjugations}).

\item We will consider the most general integrated stress tensor on a connected interval in 1+1 dimension:
\begin{align}
\label{-2}
\mathcal{O}=\int_{-b_L}^{b_R} dx \cdot [f (x) T(x) + \overline{f}(x) \overline{T}(x)]
\end{align}
\noindent and split it into two parts, $\mathcal{O}=O_L + O_R$, where
\begin{align}
\label{-3}
 O_L = \int_{-b_L}^{-a_L} dx \cdot  [f (x) T(x) + \overline{f}(x) \overline{T}(x)]      + \int_{-a_L}^{a_R} dx \cdot  [L (x) T(x) + \overline{L}(x) \overline{T}(x)]
\intertext{and}
O_R = \int^{b_R}_{a_R} dx \cdot  [f (x) T(x) + \overline{f}(x) \overline{T}(x)]      + \int_{-a_L}^{a_R} dx \cdot  [R (x) T(x) + \overline{R}(x) \overline{T}(x)]
\end{align}
\noindent both involving degrees of freedom from a common region $(-a_L, a_R)$. See figure \ref{fig:these}. Requiring the conditions on the functions
\begin{align}
\label{condi}
& R(x)+L(x) = f(x), \quad \overline{R}(x) + \overline{L}(x) = \overline{f}(x), \quad x\in (-a_L, a_R)
\\ &R(-a_L)= \overline{R}(-a_L) = L(a_R) = \overline{L}(a_R) = 0,
\end{align}

\noindent we will show that if $R$ satisfies the non-linear differential equation \eqref{-34} and $\overline{R}$ the analogous one for $\alpha \to  -\alpha$, then using the $i\epsilon$ prescription of \cite{Besken:2021oli} of comuputing commutators from the stress tensor OPE, all the nested commutators of $O_R, O_L$ are proportional to $[O_R, O_L]$, permitting a formally defined closed form Zassenhaus expansion of the type
\begin{align}
\label{-53}
e^{i \mathcal{O} t} = e^{\lambda(t)} \cdot e^{ i O_L t}  \cdot e^{i O_R t} \cdot \exp( \Omega(t) [O_R, O_L] )
\end{align}
for functions $\Omega(t)$ and $\lambda(t)$. We will see that  half-sided translations corresponding to $\mathcal{M}$ as a Rindler wedge is a special case in this class, and thus we can obtain a formal expression 
\begin{equation}
e^{i \mathcal{G} s} = e^{i G' s} e^{ i Gs} \exp( f(s)[G,G'] )
.
\end{equation}
of the type \eqref{hope}. However the involved terms won't be operators unless regularized.  (Section \ref{sec:commutator}).

\item We will show that regularizing commutators like $[G, G']$ is more tricky than regularizing $G$ and $G'$, especially when one wishes to compute {\it all} of their nested commutators. This will require use of non-analytic bump functions \cite{Lee:2012sm} for regularization of $G, G'$; these functions on the real line are infinitely differentiable smooth functions and will enable all nested commutators of $\hat{G}, \hat{G}'$ to be operators. (Section \ref{ssec:regul}).

\begin{figure}
\centering
\includegraphics[height=2.2in]{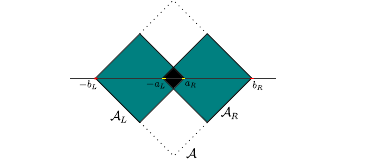}
\caption{The setup for general treatment concerning the split of $\mathcal{O}$ in \eqref{-2} into $O_L, O_R$.  If $e^{i \mathcal{O} t}$ is an operator in the algebra $\mathcal{A}$, the results can be understood as that diamond region shown in black can enable the stitching of the operators in $\mathcal{A}_L$ and $\mathcal{A}_R$ to give $e^{i \mathcal{O} t}$ in $\mathcal{A}$ through a Zassenhaus decomposition. However, one may need to regularize them over small intervals near the boundaries, which are shown in yellow and red.}
\label{fig:these}
\end{figure}

\item We will show that even after regularization of $G, G'$, the right-sided expansion of form \eqref{hope} fails when excitations spacelike to the overlap region (black diamond in figure \ref{fig:these}) are involved. The resolution of this will need a centered Zassenhaus formula,  which we will derive following the methods of \cite{Casas:2012ecz}. Consequently, we will be able to write

\begin{align}
\exp(i s \mathcal{G}) =  e^{i s \hat{G}'} & \prod_{k=2}^{\infty}  \exp\left[ (is)^{k} \left(   \frac{i^{k-2}}{k(k-2)!} D(\epsilon)   +\Sigma_k(\epsilon)\right )\right ] \cdot e^{i s \hat{G}}
 \label{product-hst}\\
\text{where }\; & D(\epsilon) = 2ai \int_{\epsilon}^{2a-\epsilon} T(x) dx  
\end{align}
\noindent for $2a = a_R, a_L = 0$ and $b_{L}, b_R \to \infty$ when comparing to figure \ref{fig:these}. $\Sigma_k(\epsilon)$ are stress tensor integrals over the intervals $(0,\epsilon)$ and $(2a-\epsilon, 2a)$ obtained by nested Wronskians of bump functions \eqref{regul}. In figure \ref{fig:these}, these intervals are shown in yellow. Further, for conjugations $e^{i s \mathcal{G}} \phi(x) e^{-i s \mathcal{G}}$ with $x$ confined to the overlap region and away from the yellow intervals, the $\Sigma_{k}$ in \eqref{product-hst} can be ignored, and the infinite product simplifies to the closed form
\begin{align}
\label{-55}
\exp(i s \mathcal{G})  \stackrel{*}{=} e^{i s \hat{G}'} \cdot \exp \left[ (e^{-s}(s+1) -1 ) D(\epsilon) \right] \cdot e^{i s \hat{G}}
\end{align}
where $*$ denotes this conditional equality. (Section \ref{sec:zas}).

\end{itemize}

In the next section, we will give a brief review of half-sided translations, and the method of \cite{Besken:2021oli} for computing commutators of stress tensors. In sections 3, 4, and 5,  we will obtain the mentioned results, and will conclude in section 6 with some future directions.

\section{Background and methods}
\label{sec:bg}
\subsection{Review of half-sided translations}

Consider a system of quantum fields in flat spacetime, in a state $|\Omega\rangle$ that is cyclic and separating for a von Neumann algebra $\mathcal{M}$ of operators. Cyclicity of the state means that $\mathcal{M}$ when acting on $|\Omega\rangle$ produces a dense subset of the full Hilbert space, and separability means that no operator in the algebra $\mathcal{M}$, except the trivial operator, annihilates $|\Omega\rangle$. These conditions are necessary and sufficient to define the Tomita operator $S$ whose polar decomposition is
\begin{align}
\label{sdeco}
&S = J_\mathcal{M} \Delta_{\mathcal{M}}^{1/2}
\intertext{where $\Delta_\mathcal{M} = \exp (- K_{\mathcal{M}})$ implements the automorphisms of the algebra $\mathcal{M}$,}
&e^{i K_{\mathcal{M}} t} \mathcal{M} e^{-i K_{\mathcal{M}} t} = \mathcal{M}, t \in \mathbb{R}, \label{modul}
\end{align}
\noindent with $J_\mathcal{M}$ an anti-unitary operator satisfying $J_\mathcal{M} \mathcal{M} J_\mathcal{M}=\mathcal{M}'$. Through the action of $J_{\mathcal{M}}$, one can see that the $S$ operator knows about $\mathcal{M}'$; in fact $\Delta_\mathcal{M}$ also generates the automorphisms of $\mathcal{M}'$. The operator $K_{\mathcal{M}}$ is unbounded in general, and the equation \eqref{modul} describes the \textit{modular flow} of $\mathcal{M}$, and the operator $K_{\mathcal{M}}$ is the \textit{modular} Hamiltonian for $\mathcal{M}$.

The conditions of existence of half-sided translations are as follows. We need two von Neumann algebra $\mathcal{M}, \mathcal{N}$ with $\mathcal{N} \subset \mathcal{M}$ that satisfy:

\begin{itemize}
\item The state $|\Omega_0\rangle$ is cyclic and separating for $\mathcal{M}$ and $\mathcal{N}$.
\end{itemize}

\begin{itemize}
\item The half-sided modular flow of $\mathcal{N}$ with $\Delta_\mathcal{M}$ lies within $\mathcal{N}$.
\end{itemize}
\begin{align}
\label{halfs}
\Delta_{\mathcal{M}}^{-it} \mathcal{N} \Delta_{\mathcal{M}}^{it} \subset \mathcal{N}, \quad t \leq 0 \text{ (or $t \geq 0$)}.
\end{align}
\noindent If satisfied, \cite{Borchers:1992cpt, Borchers:1993mis,  Wiesbrock:1993hsm, Borchers:1996hsi, Borchers:1998tvn, Borchers:1999qts}  guarantee that

\begin{itemize}
\item Half-sided translations exist and are described by a unitary group $U(s) = e^{-i \mathcal{G} s}, s \in \mathbb{R}$ generated by a positive generator $\mathcal{G}$.
\end{itemize}

\begin{itemize}
\item $U(s) |\Omega_0\rangle  = |\Omega_0\rangle, s \in \mathbb{R}$, or equivalently $\mathcal{G} |\Omega_0\rangle = 0$.
\end{itemize}

\begin{itemize}
\item $U^\dagger (s) \mathcal{M} U(s) \subset \mathcal{M}, \quad  s\leq  0$ \item $\mathcal{N} = U^\dagger (-1) \mathcal{M} U(-1)$.
\end{itemize}

\begin{itemize}
\item $[K_{\mathcal{M}}, K_{\mathcal{N}}] = - 2\pi i (K_{\mathcal{M}} - K_{\mathcal{N}})$ and $K_{\mathcal{M}}- K_{\mathcal{N}} = 2\pi \mathcal{G}$.
\end{itemize}

\begin{itemize}
\item Given an $\mathcal{M}, \mathcal{N}$ and  $|\Omega_0\rangle$ defined this way, $U(s)$ is unique.
\end{itemize}

\begin{itemize}
\item $\mathcal{M}$ needs to be a type  III$_1$ von Neumann for $\mathcal{G}$ to be non-trivial.
\end{itemize}

\noindent Let us now come to the example we will study in detail.

\begin{figure}
\centering
\includegraphics[height=2in]{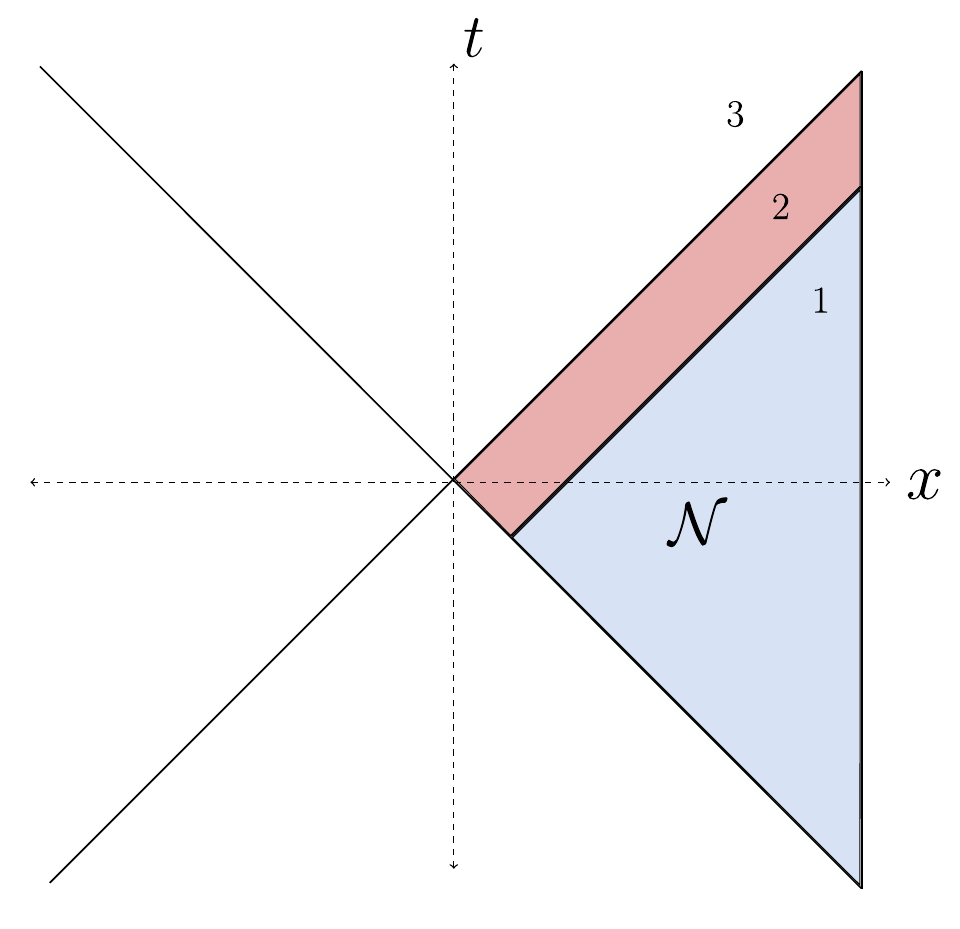}
\caption{Rindler wedges and regions 1,2,3; the condition \eqref{halfs} is satisfied by algebras of the regions 1 and 2.}
\label{fig:rindl}
\end{figure}

\subsection{Half-sided translations for Rindler wedge}

\noindent Consider that our conformal system is in the vacuum state $|0\rangle$ corresponding to time coordinate $t$, and let the algebras $\mathcal{M}, \mathcal{N}$ correspond to the causal diamond regions
\begin{align}
\label{mandn}
\Diamond_M &= \{x^{+} > t_0 + x_0,\quad x^{-} < t_0 - x_0\},\quad \Diamond_{N} = \{ x^{+}> t_0 + x_0, x^{-} < t_0 - x_0- 2a\}.
\end{align}
\noindent These are Rindler wedges, with $M=(x_0,\infty)$ at $t=t_0$, and $N= (x_0+a, \infty)$  at $t=t_0- a$.  The vacuum state $|0\rangle$ is cyclic and separating for the algebras $\mathcal{M}, \mathcal{N}$, which obey $\mathcal{N} \subset \mathcal{M}$ and satisfy the property \eqref{halfs}. Thus half-sided translations exist and are generated by $\mathcal{G}=\frac{1}{2\pi} (K_{\mathcal{M}}-K_{\mathcal{N}})$,  where  $K_{\mathcal{M}}$ and $K_{\mathcal{N}}$ are $2\pi$ times the boost operators preserving $(t,x) = (t_0,x_0)$ and $(t,x)=(t_0 - a,x_0+ a )$ respectively \cite{Witten:2018zxz, Cardy:2016fqc, Callebaut:2018xfu}. This result is in fact independent of $(x_0, t_0)$,  and is given by
\begin{align}
\label{resul}
\mathcal{G} = 2a P^{+} = a (H+P)  = 2a \int_{-\infty}^{\infty} T(x) dx.
\end{align}
\noindent It is derived using a Poincar\'e generator and a stress tensor method in \ref{diffe} and the result allows us to set $x_0=0, t_0=0$. See figure \ref{fig:rindl}, where the Minkowski space is divded into three regions, where $R_1 = \Diamond_\mathcal{N}, R_{2} \cup R_1= \Diamond_{\mathcal{M}}$ and complement of $R_1 \cup R_2$ is $R_3$. We will refer to these regions later.

Each $K$ decomposes in terms of entanglement Hamiltonians $H_{\mathcal{M}}, H_{\mathcal{N}}, H_{\mathcal{M}'}$,  and $\mathcal{H}_{\mathcal{N}'}$ as discussed in the introduction, for
\begin{align}
\label{entan}
H_\mathcal{M} &= 2 \pi \int_{0}^{\infty} dx\; x \cdot T_{00}(x,0),\\ H_{\mathcal{M}'} &= - 2\pi \int_{-\infty}^{0} dx\; x \cdot T_{00}(x,0)
\\H_\mathcal{N} &= 2 \pi \int_{a}^{\infty} dx\; (x-a) T_{00}(x,  - a), \\ H_{\mathcal{N}'}&=  - 2 \pi\int_{-\infty}^{a} dx\; (x-a) T_{00}(x,- a).
\end{align}
\noindent Let us now take the first step to compute the commutators of these quantities, by studying commutators of stress tensors in 2d CFT.

\subsection{Commutators from the OPE}

We will now address how to compute commutators of the stress tensor from their OPE. This can be done via the $i\epsilon$ prescription \cite{Besken:2021oli}. Using the Wick rotation $\tau = i t$ or $t = -i \tau$,  we will apply the rule
\begin{align}
\label{theru}
&[O_i (t, x), O_j(0)] = \lim_{\epsilon \to 0}  [O_i(t-i\epsilon, x) O_j(0) - O_i (t+i\epsilon, x) O_j].
\end{align}
Note that stress tensor OPEs differ from the usual CFT definition by factors of $-2\pi$, which is explained in \cite{Polchinski:1998rq}. The $T(z)T(w)$ OPE is then\begin{align}
\label{theop}
&T(z) T(w) = \frac{c}{8\pi^2 (z-w)^{4}} - \frac{T(w)}{\pi (z-w)^2} - \frac{\partial T(w)}{2 \pi (z-w)}.
\end{align}To get a useful answer from \eqref{theru}, we need to integrate both sides with respect to test functions. But first we must write \eqref{theop} in terms of $t$. We will be using $z=x-t$ as our variable, therefore,\begin{align}
\label{rulea}
 &[T(z), T(w)] = \lim_{\epsilon \to 0} [T(x-(t-i\epsilon)) T(w)  - T(x-(t+i\epsilon)) T(w)]
\intertext{which after a contour integral against a test function (either in upper half or the lower half plane) gives}
&[T(z), T(w)]= -i (T(z) + T(w)) \cdot \partial_z\delta(z-w) + \frac{i c}{24 \pi} \partial_z^3 \delta(z-w) \label{TT}.
\end{align}Note the very important minus sign compared to \cite{Besken:2021oli} that arises from our convention of writing $z=x+i\tau=x-t$. A similar calculation can be done to obtain $[T(z), \phi(w,\overline{w})]$ for primary field $\phi(w,\overline{w})$ of conformal weight $(h, \overline{h})$ that gives
\begin{align}
\label{t-phi}
[T(z), \phi(w, \overline{w})] &= -i\left( \phi(z,\overline{w}) + (h-1)\phi(w, \overline{w})\right)  \cdot \delta'(z-w).
\end{align}
Note that when dealing with anti-holomorphic components a minus sign will be picked because $\overline{z} = x + t$ and \eqref{rulea} is sensitive to the sign of $t$. Further note that because $T(x) \overline{T}(y)$ OPE does not have any singular terms, their commutator is zero. An explicit calculation of unequal time commutator relations can be done easily for the free massless scalar field to verify these statements. We will use these observations in section \ref{sec:commutator}.

\section{On $G$ and $G'$ as operators}

\label{sec:g-g'}

Quantum fields are postulated to be operator valued distributions, which are symmetric bilinear forms over a dense domain $D \subset \mathcal{H}$ \cite{Haag:1962pqft}. This means a field $A(x)$ has well-defined matrix elements $\langle \Psi | A(x)| \Phi \rangle$ for $\Psi, \Phi \in D$ but $A(x)$ is not an operator itself.  The products of fields, or their derivatives, are similarly operator valued distributions \cite{Haag:1996hvx}, although one may need to regularize them first. For instance, the product of two fields e.g. $A(x) A(y)$ is divergent as $x\to y$. This is well known in CFT, where one can do a normal ordering when defining composite fields such as the stress tensor \cite{GinspargCFTLectures}. Consequently, composite fields can be defined to have finite matrix elements. However, bilinear forms are generally not operators.

Producing operators from fields requires smearing against smooth functions \cite{Haag:1962pqft,Witten:2018zxz}. These functions are usually required to have fast decay at infinity to get sensible operators, but this isn't always the case. For example, the boost $K_\mathcal{M}$ associated to the Rindler wedge $\Diamond_{\mathcal{M}}$ \eqref{mandn}
\begin{align}
\label{boost}
K_{\mathcal{M}} = \int_{-\infty}^{\infty} x \, T_{00}(x) \, dx
\end{align}
\noindent is a well-defined operator even though the smearing function $x$ is unbounded at infinity. As we have been discussing,  
this property is not enjoyed by the entanglement Hamiltonians $H_{\mathcal{M}}$ and $H_{\mathcal{M}'}$ given in \eqref{entan}, even though $K_{\mathcal{M}} = H_{\mathcal{M}} - H_{\mathcal{M}'}$. While $H_{\mathcal{M}}, H_{\mathcal{M}'}$  are well-defined bilinear forms, they fail to be operators \cite{Witten:2018zxz}. Below we will show this and study the reason for this failure, which will be important later for discussion of $G$ and $G'$.

\subsection{Entanglement Hamiltonians as operators?}

\noindent Let us show that $H_\mathcal{M}$ and $H_{\mathcal{M}'}$ are not operators by a contradiction proof. We will use the fact that $K$ is a well-defined operator with the vacuum $|0\rangle$ in its domain, and the OPE of stress tensors to arrive at our result.

Let us start by assuming that $H_{\mathcal{M}}$ and $H_{\mathcal{M}'}$ are densely defined operators on domains $D_{\mathcal{M}}$ and $D_{\mathcal{M}'}$ and hence have unique adjoints. As $H_\mathcal{M}, H_{\mathcal{M}'}$ are integrals of $T_{00}(x)$, they are symmetric operators. Now, note that the domain of $K = H_{\mathcal{M}}- H_{\mathcal{M}'}$ contains $|0\rangle$. In fact $K|0\rangle=0$, as boost is a symmetry. Crucially,  $D_{K} \subset D_{\mathcal{M}} \cap  D_{\mathcal{M}'}$,  and therefore,
\begin{align}
\label{domai}
 |0\rangle \in D_K \implies |0\rangle \in D_{\mathcal{M}}, D_{\mathcal{M}'}.
\end{align}
\noindent In other words, $H_{\mathcal{M}}|0\rangle = H_{\mathcal{M}'}|0\rangle$ is a state of finite norm. This we will show is false, by computing the norm using the OPE. Because $H_{\mathcal{M}}$ is symmetric and $|0\rangle\in D_{\mathcal{M}}$, note that norm squared of $H_{\mathcal{M}}|0\rangle$ is $\langle 0 | H_{\mathcal{M}}^2| 0 \rangle$ which is\footnote{ As the form is unbounded, it follows that $| \chi\rangle, H_\mathcal{M}|\chi \rangle \in D_{\mathcal{M}}$, then $\langle H_{\mathcal{M}} \chi |  H_{\mathcal{M}} \chi\rangle  = \langle \chi | H_M^2| \chi \rangle$. Last step is due to the fact that a symmetric operator $A$ coincides with its adjoint $A^\dagger$ on the domain of $A$. }
\begin{align}
\label{entan-2}
&||H_{\mathcal{M}} | 0 \rangle ||^2 = \int_0^{\infty} dx \int_0^{\infty} dy  \cdot xy \cdot \left\langle 0| T_{00}(x+i\epsilon) T_{00}(y)|0\right\rangle
\end{align}
\noindent where we have put in an $i\epsilon$ to appropriately define the product $H_{\mathcal{M}}^2$; we will it take to zero at the end of calculation. Writing $T_{00}(x,0)=T(x) + \overline{T}(x)$, and using the OPE, the central charge term gives
\begin{align}
   \frac{c}{4 \pi^2} \int_0^{\infty} dx \int_0^{\infty} dy  \cdot xy \cdot \frac{1}{(x-y+i\epsilon)^{4}}
= \frac{c}{24 \pi^2} \int_0^{\infty}  \frac{x dx}{(x+i\epsilon)^2} \label{log-div}
\end{align}
\noindent which is a divergent integral, both because of $x=0$ contributions and the asymptotic contributions. Therefore $|0\rangle \notin D_{\mathcal{M}}$. We could have similarly done it for $H_{\mathcal{M}}'$. Thus we have a contradiction, and therefore $H_{\mathcal{M}}, H_{\mathcal{M}'}$ are not operators. $\square$

\subsubsection{Extension using Reeh-Schlieder theorem}

Concerning the norm of states $H_{M} |\chi\rangle$,  we can actually derive a stronger result. It was noted in \cite{Witten:2018zxz} that norm of $H_{\mathcal{M}} |\chi\rangle, H_{\mathcal{M}'}|\chi\rangle$ has a universal ultraviolet divergence. Using the Reeh-Schlieder theorem and the OPE of the stress tensor we can explicitly show that $H_{\mathcal{M}}|\chi\rangle, H_{\mathcal{M}} |\chi\rangle$ norms are infinite for $\chi$ in an arbitrarily large dense set of the vacuum sector of the Hilbert space, with both ultraviolet and infrared divergences.

The key point in the derivation concerns the radius of convergence of the stress tensor OPE, which is determined by the distance to the nearest operator not included in the expansion \cite{Polchinski:1998rq}. If we identify some states $\chi$ such that in computation of $||H_\mathcal{M} |\chi\rangle||^2$ we can draw a circle around $T(x)$ and $T(y)$ containing no other insertion as $x, y \to 0$ and $x,y\to \infty$, then their norm squares are of the type \eqref{log-div}. 

The Reeh-Schlieder theorem tells us that such states $\chi$ can approximate any state in the vacuum sector of the Hilbert space $\mathcal{H}_0$. More precisely, let $\mathcal{A}_\mathcal{U}$ be the polynomial algebra associated to an open region $\mathcal{U}$ in spacetime, that is, algebra generated using operators of the form $\int_{\mathcal{U}} f(x) \phi(x) dx$. The theorem states that $\mathcal{A}_{\mathcal{U}} |0\rangle$ is a dense subset of $\mathcal{H}_0$ for any $\mathcal{U}$ \cite{Haag:1996hvx, Witten:2018zxz}.

Now, for us, provided that $\mathcal{U}$ is a bounded region and does not contain $(0,0)$, we can always find an $\alpha_{\mathcal{U}}, \beta_{\mathcal{U}}$ such that for $0\leq x,y < \alpha_{\mathcal{U}}$ and $x,y \geq \beta_{\mathcal{U}}$, the OPE for $T(x)T(y)$ converges. For any such $\mathcal{U}$, we can take any $\chi \in \mathcal{A}_{\mathcal{U}}|0\rangle$, and we will find that $||H_{\mathcal{M}} |\chi\rangle||^2$ diverges because of the $\chi$ independent central charge term. For instance, the asymptotic contribution is 
\begin{align}
\label{diver-2}
\int_{\beta_{\mathcal{U}}}^{\infty} dx \int_{\beta_{\mathcal{U}}}^{\infty} dy  \cdot xy \cdot \frac{1}{(x-y+i\epsilon)^{4}} =  \int_{\beta_{\mathcal{U}}}^{\infty}  dx \; x \frac{3 \beta_{\mathcal{U}}+i \epsilon -x}{6 (\beta_{\mathcal{U}}+i \epsilon -x)^3}  \sim  \int_{\beta_{\mathcal{U}}}^{\infty} \frac{dx}{x}
.
\end{align}
\noindent See figure \ref{fig:rindl} for an illustration.  Similarly there is a contribution from near $x=0$.

Let us note two key points in the above discussion which will be important below. The first is that the integrals \eqref{log-div}, \eqref{diver-2} are divergent not only because of $x = 0$ contributions, but also due to infrared contributions coming from the slow decay of $1 /x$. The second point is the observation that these integrals are marginally divergent. That is, if the integrand had been changed from $x^{-1} \to x^{-1+\kappa}$ near the origin, and $x^{-1} \to x^{-1+\kappa}$ asymptotically, for any $\kappa > 0$, the integrals would have converged. We will next see that $||G|\chi\rangle||^2$ and $||G'|\chi\rangle||^2$ are not infrared divergent, and their UV divergence can be regularized conveniently because of its marginal nature. 

\begin{figure}
\centering
\includegraphics[height=2in]{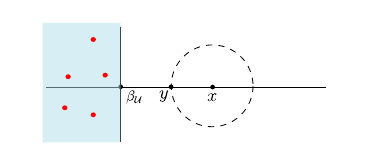}
\caption{The OPE of two stress tensor insertions $T(x) T(y)$ converges for $x, y > \beta_{\mathcal{U}}$ where $\beta_{\mathcal{U}}$ is chosen such that all excitations to produce $\chi$ lie before $x<\beta_\mathcal{U}$.}
\label{fig:theop}
\end{figure}

\subsection{Forms $G,G'$ to operators $\hat{G},\hat{G}'$}

\noindent Let us rewrite $G = H_{\mathcal{M}} - H_{\mathcal{N}}, G' = H_{\mathcal{N}'}- H_{\mathcal{M}'}$ obtained from \eqref{entan} using properties of the stress tensor in 2d CFT as follows. From the definition,
\begin{align}
\label{ggspl}
G &= \frac{H_M - H_N}{2\pi}  = \int_0^{\infty} dx \cdot x \cdot T_{00}(x,0) - \int_a^{\infty}dx \cdot (x-a) \cdot T_{00}(x,-a)
\\&= \int_0^{\infty} dx \cdot x \cdot T_{00}(x,0) - \int_0^{\infty}dx \cdot x \cdot T_{00}(x+a,-a).
\intertext{Shifting the integration variable, and using the transformation rule $T_{00}(x,t)=T(x-t)+\overline{T}(x+t)$,}
G &= \int_0^{\infty} dx \cdot x \cdot [T(x)+\overline{T}(x)] - \int_0^{\infty}dx \cdot x \cdot [T(x+2a)+\overline{T}(x)]
\\ &= \int_0^{\infty} dx \cdot x \cdot [T(x)-T(x+2a)]
\\ &= \int_0^{2a} dx \cdot x \cdot T(x)+ 2a\int_{2a} ^{\infty}dx  \cdot T(x) \label{Gfinal}.
\intertext{Likewise,}
&G' = \int_{0}^{2a} dx \cdot (2a-x) \cdot T(x) + 2a \int_{-\infty}^{0} dx \cdot T(x) \label{G'final}
\end{align}
\noindent These mimic $2a P^{+}$ as $x \to \pm \infty$, which is a well-defined operator, and therefore the norm of $G|\chi\rangle, G'|\chi\rangle$ cannot have a universal infrared divergences. Near $x=0$ however, the behaviour is same as $H_{\mathcal{M}}$,  and therefore the norms of the states $G |\chi\rangle, G |\chi'\rangle$ show a universal UV divergence. However, as we saw in the last subsection, this divergence is marginal and can be fixed conveniently.

Let us replace $G, G'$ by suitable $\hat{G}, \hat{G}'$. It is important to recognize that a change in $G$ must be balanced with a change in $G'$ to maintain $\mathcal{G} = G + G' = \hat{G} + \hat{G}'$, and because there are two points requiring regulation, i.e. $x=0$ and $x=2a$,  we will need two functions $\mu$ and $\nu$. Let
\begin{align}
\label{hatg}
& \hat{G}= 2a\int_{2a} ^{\infty}dx  \cdot T(x)  + \int_0^{\epsilon} \mu(x) T(x) +   \int_\epsilon^{2a-\epsilon} dx \cdot x \cdot T(x)+  \int_{2a-\epsilon}^{2a} \nu(x) T(x)
\\ & \hat{G}'= 2a\int^{0} _{-\infty}dx  \cdot T(x)  + \int_0^{\epsilon} [ 2a-\mu(x) ] T(x) +   \int_\epsilon^{2a-\epsilon} dx \cdot (2a-x) \cdot T(x)+  \int_{2a-\epsilon}^{2a} [ 2a-\nu(x) ] T(x)
\end{align}
\noindent for $\epsilon \ll 1$. One possible choice for $\mu(x)$ and $\nu(x)$ is the minimal regularization
\begin{align}
\label{minim}
\mu(x) &=x\left(\frac{x}{\epsilon}\right)^{\epsilon},\quad \nu(x) = 2a-  \frac{1}{\epsilon^{\epsilon}} (2a-x)^{1+\epsilon}
\end{align}
\noindent which is sufficient to ensure that norms of $\hat{G}|\chi\rangle, \hat{G}'|\chi\rangle$ do not display universal UV divergences. This prescription satisfies continuity at $x=\epsilon, 2a-\epsilon$, which is an obvious requirement.  Note that in the limit $\epsilon \to 0$, $\mu (x) = \nu(x) = x$, which suggests $G,G'$ by themselves possibly fall just marginally short of being operators. If the divergence was stronger, requiring a different regularization like $\mu(x) \sim x^2$, continuity would demand $\frac{1}{\epsilon}$ factors in the denominator which would diverge in the limit $\epsilon \to 0$.

Let us note two more points in this subsection. First is that the contributions from the additional boundaries introduced in the various integral splits we considered cancel against each other, therefore we don't need to regularize integrands there. Second point, which we will return to in section \ref{ssec:regul}, is that any analytic function used for regularization will have a derivative problem at the points $x=\epsilon, x=2a-\epsilon$. For example, $\mu(x) = x^{1+\epsilon} \epsilon^{-\epsilon}$ at $x=\epsilon$ is $\epsilon$, but its derivative at $x=\epsilon$ is not zero. We can regularize using higher degree polynomials which would have $N$ derivatives continuous at $x=\epsilon$, but $N$ cannot be infinite. This will be an important point when studying commutators later, which will force us to use non-analytic bump functions for regulation.

\subsection{The conjugations $e^{iGs} \phi e^{-i Gs}$ and $e^{iG's} \phi e^{-i G's}$}
\label{sec:conjugations}

\noindent What is the nature of the formal operators $G, G'$, and what dynamics do they generate? Is it consistent with causality? To understand this, let us study the behaviour of $e^{iGs}$ and $e^{i G's}$. At the end, we will see that our discussions will allow conclusions for $ e^{is\hat{G}}, e^{is\hat{G}'}$. Let us start with the quantity
\begin{align}
\label{eigs}
\Phi(s; x,t) &= e^{iGs} \phi(x,t) e^{-i Gs}
\intertext{for a primary field $\phi$. This will be a bilinear form again, because $G$ is known to have problems being an operator as discussed above.  However we can do this computation anyway by the Hadamard lemma,}
 \Phi(s) &= \sum_{n=0}^{\infty} \frac{(is)^n}{n!} \text{ad}_G^n \phi(x,t)
\end{align}
where the notation is $\text{ad}_G \phi = [G,\phi], \text{ad}_G^k \phi = [G, \text{ad}^{k-1} _G \phi]$, and $\text{ad}_G^0 \phi = \phi $. Now $[G,\phi]$ is given by\footnote{More precisely these integrals should be understood as integrals over matrix elements of the involved operator valued distributions. }
\begin{align}
[G, \phi(x,t)] &= \int_{0}^{2a} dz \cdot z [T(z), \phi(x,t)] + 2a \int_{2a}^{\infty} dz \cdot [T(z), \phi(x,t)].
\intertext{This requires us to obtain $[T(z), \phi(x,t)]$ which can be done using $i\epsilon$ prescription starting from the OPE. We will find \eqref{t-phi}:}
[T(z), \phi(w, \overline{w})] &= -i\left( \phi(z,\overline{w}) + (h-1)\phi(w, \overline{w})\right)  \cdot \delta'(z-w) \nonumber
\intertext{where $\overline{w} = x+t, w=x-t$ and $h$ is the holomorphic conformal weight for $\phi$. Let us at first place $(x,t)$ in region $R_1 = \Diamond_{\mathcal{N}}$ of figure \ref{fig:rindl} and define for convenience}
&I_1 = 2a \int_{2a}^{\infty} dx \cdot T(x), \quad I_2 = \int_0^{2a} dx \cdot x\cdot  T(x)
\intertext{where the notation references the same figure. Then}
[G, \phi(x,t)]&= 2a\int_{2a}^{\infty} dz \cdot [T(z), \phi(x,t)]
\\& =- i\, 2a\int_{2a}^{\infty} dz \cdot \left( \phi(z) + (h-1)\phi(w, \overline{w})\right) \cdot  \delta'(z-w) 
\\ & = 2a i\int_{2a}^{\infty} dz \cdot \partial_z\left( \phi(z) + (h-1)\phi(w, \overline{w})\right) \cdot  \delta(z-w) 
\intertext{provided we are far away from boundary $x-t=2a$. This gives us $ i\partial_w\phi(w)$,  but $w=x-t=-x^{-}$, and hence}
[G, \phi(x,t)]&= - 2a i\partial_- \phi(x,t) \quad \text{or} \quad \frac{1}{i}[\phi(x,t), G]= 2a \cdot \partial_- \phi(x,t) \label{reg1}
\intertext{which is the standard Heisenberg equation of motion for $\phi$ under the operator $G$.   Due to the constant velocity, $\Phi(s)$ becomes a standard Taylor series, describing the flow of $\phi$ along the null line $x^{-}$.  This establishes that $G$ is acting like $\mathcal{G}$ provided we are in region 1.  When we approach the boundary, we have to be more careful to include the boundary term. Then,}
[I_1, \phi]  = -  2ai &[\phi(z,\overline{w}) + (h-1) \phi(w,\overline{w})]\delta(z-w)\Big|^{\infty}_{2a} + 2a i \int_{2a}^{\infty} dz \;\partial_z \phi(z) \delta(z-w)
\intertext{which at the boundary $w=2a$ becomes}
& a i \cdot \partial \phi(2a,\overline{w}) + 2a i \frac{1}{2\epsilon} \cdot [\phi(2a,\overline{w})+ (h-1) \phi(2a,\overline{w})]
\intertext{by approximating $\delta$ function as $\frac{1}{2\epsilon}$ on a width of $2\epsilon$. This is important because integration by parts requires the boundary term to be regular. On the other hand, for $[I_2, \phi]$}
 [I_2, \phi]  =  \int_{\epsilon}^{2a}& dz \cdot  z \cdot [T(x), \phi(y)]
\\  = -i  \int_{0}^{2a} dz & \cdot  z \cdot ( \phi(z,\overline{w}) + (h-1) \phi(w,\overline{w}) ) \cdot \delta'(z-w)
\\ = -iz [\phi(z,\overline{w})+ &(h-1) \phi(w,\overline{w})] \cdot \delta(z-w)\Big|^{2a}_{0}  + i \int_{0}^{2a} dz \cdot [z \partial\phi(z,\overline{w})+h\phi(w,\overline{w}) ] \delta(z-w).
\intertext{At the boundary $w=2a$, this term also contributes, and we get}
- 2a i [\phi(2a,\overline{w}) &+(h-1) \phi(2a, \overline{w})]\cdot \frac{1}{2\epsilon}+ \frac{i}{2} \left( 2a \partial\phi(2a, \overline{w}) + h \phi(2a, w)\right).
\intertext{Hence when $I_1, I_2$ are combined, the singularity cancels for $w=2a$, but note there still is a discontinuity because of  $\frac{h}{2}\phi(2a,\overline{w})$. This is due to the fact that writing $G = \int_0 ^{\infty} dx \cdot f(x) T(x)$, the function $f$ has a discontinuous derivative at $x=2a$. For $w<2a$, only $I_2$' s integral contributes,}
[I_2, \phi(w,\overline{w})] &= i \cdot w \partial_w \phi(w,\overline{w})  + i h \phi(w,\overline{w})
\intertext{where the first term produces dilation, and the second produces a multiplicative effect. We can see this better in $\Phi(s)$.  Take the double commutator,  }
 [I_2, [I_2, \phi(w,\overline{w})]] &= i \cdot w [I_2, \partial\phi(w,\overline{w})]   + ih \cdot  [I_2, \phi(w, \overline{w})].
\intertext{We can take the derivative action on $w$ outside the commutator, to write}
[I_2,[I_2, \phi]]& = i(h +w \partial)[I_2, \phi(w,\overline{w})]
\\ & = i(h +w \partial)[i(h+w \partial)\phi(w,\overline{w})].
\intertext{We therefore can see,}
& \text{ad}_{I_2}^{k} \phi(w,\overline{w}) =i^{k}(h+w\partial)^{k}\phi(w,\overline{w})
\intertext{and thus, if $(w,\overline{w})$ was in region 2,}
\Phi(s) &= \sum_{j=0}^{\infty} \frac{(is)^{j}}{j!}  \text{ad}_{I_2}^{j} \phi(w,\overline{w})
\\  &= \sum_{j=0}^{\infty} \frac{(-s)^j}{j!}  (h+w\partial)^{j}\phi(w,\overline{w})
\\ &= e^{-s (h+w \partial) }\phi(w,\overline{w}) = e^{-sh} \phi(e^{-s}w, \overline{w}). \label{scale-dim}
\end{align}
\noindent Thus we see, that operator is moving towards $w=0$ (through a dilation), the boundary of region 2 and 3, and its also diminishing due to the $e^{-sh}$ factor.

\subsubsection*{Calculations for $G'$}

Similar calculations can be done for $e^{iG's}$ by symmetry. We will find
\begin{align}
\label{leftt}
& [G', \phi(w,\overline{w})] = i (2a -w ) \partial_w \phi(w,\overline{w}) -ih \phi(w,\overline{w})
\intertext{for $(w,\overline{w})$ in region 2, which implies}
& \Phi(s) = e^{-s(-h + (2a-w)\partial)}\phi(w,\overline{w}) = \exp[ (sh + s(w-2a)\partial) ]\phi(w,\overline{w}).
\intertext{as long as $(w(s), \overline{w})$ remains in region 2. By a change of variable, $u = \ln (w-2a)$,  the exponential operator can become a translation operator which simplifies to}
\Phi(s)& =  e^{sh} \phi(-e^{s} (2a-w) + 2a, \overline{w})
\intertext{and we can find the time till this expression is valid by setting $e^{s} (2a-w) + 2a = 0$, corresponding to $w=0$ boundary of region 2, which gives }
& s_{1} = \ln \frac{2a}{2a - w}
\end{align}
\noindent If $(w,\overline{w})$ was in region 3, then $G'$ would act like $\mathcal{G}$, just like how $G$ did in \eqref{reg1}.

Now what if we replace $G,G'$ by $\hat{G}, \hat{G}'$? The only changes will be near the boundaries. However, the factors $e^{-sh}, e^{sh}$ will persist. Conjugation by $e^{i \hat{G}s}$ or $e^{i \hat{G}'s}$ therefore will change the operator norm, and thus these operators cannot be unitary. In other words, while $\hat{G}, \hat{G}'$ may be symmetric operators, they cannot be self-adjoint.\footnote{In the theory of bilinear forms, this corresponds to the case that $\hat{G}, \hat{G}'$ as forms are not closable \cite{Reed:1981fa}.}

\noindent $ $

\section{Commutator computations}
\label{sec:commutator}

We will now compute the commutators for a general class of integrated stress tensors, and later specialize to the half-sided translations. The commutator of stress tensor is obtained by taking products of fields without regularizing them, and therefore $\delta(x)$ singularities appear naturally in them. As we will be integrating over these singularities, we will need to be careful especially in integration by parts.  As mentioned above, for using the formula of integration by parts, we will need the boundary terms to be regular, and this will require us to express delta function as a $1/2\epsilon$ on interval of width $2\epsilon$,  which we will take to zero at the end.

\subsection{Commutators of integrals of stress tensor}

Let us consider the operator
\begin{align}
\label{-8}
\mathcal{O}=\int_{-b_L}^{b_R} dx \cdot [f (x) T(x) + \overline{f}(x) \overline{T}(x)].
\end{align}
\noindent We will split it into two parts, $\mathcal{O}=O_L + O_R$, where
\begin{align}
\label{-9}
O_L = \int_{-b_L}^{-a_L} dx \cdot  [f (x) T(x) + \overline{f}(x) \overline{T}(x)]      + \int_{-a_L}^{a_R} dx \cdot  [L (x) T(x) + \overline{L}(x) \overline{T}(x)]
\intertext{and}
O_R = \int^{b_R}_{a_R} dx \cdot  [f (x) T(x) + \overline{f}(x) \overline{T}(x)]      + \int_{-a_L}^{a_R} dx \cdot  [R (x) T(x) + \overline{R}(x) \overline{T}(x)].
\end{align}
\noindent The condition $O_L+O_R =\mathcal{O}$ constrains the functions $L,R, \overline{L}$ and $\overline{R}$:
\begin{align}
\label{-10}
R(x)+L(x) = f(x), \quad \overline{R}(x) + \overline{L}(x) = \overline{f}(x), \quad x\in (-a_L, a_R).
\end{align}
\noindent Further, we would like to require continuity conditions, that
\begin{align}
\label{-11}
R(-a_L)= \overline{R}(-a_L) = L(a_R) = \overline{L}(a_R) = 0.
\end{align}
\noindent Using causality, these conditions allow us to write $[O_R, O_L]$ as integrals over just $(-a_L, a_R)$:
\begin{align}
\label{-12}
[O_R,O_L] = \int_{-a_L}^{a_R} dx \int_{-a_L}^{a_R} dy \left( R(x)L(y) [T(x),T(y)] + \overline{R}(x) \overline{L}(y) [\overline{T}(x) , \overline{T}(y)]\right).
\end{align}
\noindent Below at first we will set $\overline{f}(x) =\overline{R}(x) =\overline{L}(x) =0$ which will be sufficient to discuss the special case of half-sided translations for Rindler wedge. As the OPE of $T$ and $\overline{T}$ does not have any singular terms, $[T(x), \overline{T}(y)]=0$, therefore the extension to the more general case can be done in a straightforward manner at the end. Also, we will set $I = (-a_L, a_R)$ for brevity. Now we can write using \eqref{theop}
\begin{align}
\label{-13}
[O_R,O_L]=   \int_I dy \int_I dx \cdot L(y) R(x)\left(   -i [T(x) + T(y)] \cdot  \delta'(x-y) +   \frac{ic}{24 \pi}       \cdot  \delta'''(x-y)  \right).
\end{align}
\noindent Thus,

\begin{align}
\label{-14}
% [O_R, O_L] &= i \int_I dy \int_I dx \cdot L(y)  \partial_x( R(x) \cdot [T(x) + T(y)] )   
% \\ \nonumber - &   \frac{c}{24 \pi} \cdot  \cdot  R'''(x)  \delta(x-y) + \Sigma_1
[O_R, O_L]= i \int_I dy \int_I dx \, L(y) \left( \partial_x( R(x) \cdot [T(x) + T(y)] )   -   \frac{c}{24 \pi}  R'''(x) \right) \delta(x-y) + \Sigma_1
\end{align}

\noindent where
\begin{align}
\label{-15}
\Sigma_1 = \int_I dy \left(  - i L(y) R(x)\cdot  [T(x)+T(y)] \delta(x-y)\Big|_{-a_L}^{a_R} + \frac{i c L(y)R(x)}{24 \pi}  \cdot \delta''(x-y)\Big|^{a_R}_{-a_L}     \right).
\end{align}
\noindent Because of the boundary condition $R(a_{-L})=0$, this becomes
\begin{align}
\label{-16}
\Sigma_1 = \int_I dy \left(  - i L(y) R(a_R)\cdot  [T(a_R)+T(y)] \delta(a_R-y) + \frac{i c L(y)R(a_R)}{24 \pi}  \cdot \delta''(a_R-y)  \right).
\end{align}
\noindent Because the integral is over the interval $(-a_L,a_R)$ and there is a $\delta$ function at the boundary, such an evaluation would need a lot of care, but in this case since $L(a_R)=0$, this integral is easily zero. Therefore
\begin{align}
\label{-17}
&[O_R, O_L]= i \int_I dy \int_I dx \cdot L(y) \left( R'(x) \cdot [T(x)+T(y)] + R(x) \cdot T'(x)   -   \frac{c}{24 \pi}    \cdot R'''(x)\right)\delta(x-y)
\\ & = i \int_I dy   \cdot L(y) \left( 2 R'(y) \cdot T(y) + R(y) \cdot T'(y)   -   \frac{c}{24 \pi}    \cdot R'''(y)\right).
\end{align}
\noindent The non-trivial operator content comes from first integral, which is
\begin{align}
\label{-18}
i \int_I dy \; L(y) \cdot [2R'(y) \cdot T(y) + R(y) \cdot T'(y)].
\end{align}
\noindent Integration by parts of the second term gives
\begin{align}
\label{-19}
 i \left( L(y) R(y) \cdot T(y )\Big|^{a_R}_{-a_L} - \int_I dy  \cdot T(y) \cdot [R'(y) L(y) +L'(y)R(y)] \right).
\end{align}
\noindent This new boundary term here is easily seen to be zero. Thus
\begin{align}
\label{-20}
[O_R, O_L] = -\frac{ic}{24 \pi} \int_I dy \cdot L(y) R'''(y) + i \int_I dy \cdot \left[  L(y) R'(y)  -L'(y) R(y) \right ] T(y).
\end{align}
\noindent If $f$ is a constant, $L'(y)=-R'(y)$, and the non-trivial bulk term becomes $i f\int_I dy \cdot R'(y) T(y)$. At this stage let us note the appearance of Wronskian of functions $L, R$ defined by $W(L,R)=L R'-L'R$. We will see that these appear each time we take a nested commutator. Let us define $\omega(x)=L(x)R'(x)-L'(x)R(x)$ below.

\subsection{Nested commutators}

Next, let us evaluate  $[O_R,[O_R,O_L]]$. Because the central charge term is proportional to identity operator,
\begin{align}
\label{neste}
 &[O_R,[O_R,O_L]] = i \int_I dy  \int_I dx  \cdot [R(x)T(x), \;\; \omega(y) T(y)]
\\ = i\int_I dy \int_I dx &   \cdot \omega(y) \cdot R(x) \cdot  \left( -i[T(x)+  T(y)] \delta'(x-y)   +  \frac{ic}{24 \pi} \delta'''(x-y)  \right).
\end{align}
\noindent This is the same form as $[O_R, O_L]$ with $L \to \omega$. Therefore
\begin{align}
\label{-22}
 [O_R, [O_R, O_L]]= -\frac{i^2c}{24 \pi} \int_I dy \cdot \omega(y) R'''(y) + i^2 \int_I dy \cdot [\omega(y) R'(y)-\omega'(y)R(y)]T(y) + \Sigma
\end{align}
\noindent but this time $\Sigma$ is not zero trivially because $\omega(a_R)=-L'(a_R) R(a_R)$.  Its evaluation will require a careful treatment of the integrals. For this we can approximate $\delta$ function as $\frac{1}{2\epsilon}$ over the interval $(a_R -\epsilon,a_R+\epsilon)$ for $\epsilon \to 0$. Thus
\begin{align}
\label{-23}
&\Sigma = -i R(a_R)^2 L'(a_R) T(a_R) +R(a_R) \int_{a_R-\epsilon}^{a_R} dy \; \omega(y) \left(  - i  \cdot  [T(a_R)+T(y)] \frac{1}{2\epsilon} + \frac{i c \delta''(a_R-y)}{24 \pi}       \right)
\\ &= -i R(a_R)^2 L'(a_R) T(a_R) - \frac{i}{2}R(a_R)  \; \omega(a_R)   \cdot  2T(a_R) +\int_{a_R-\epsilon}^{a_R} dy \cdot \omega(y) \cdot  \frac{i c \delta''(a_R-y)}{24 \pi}
\\ &= \int_{a_R-\epsilon}^{a_R} dy \cdot \omega(y) \cdot  \frac{i c \delta''(a_R-y)}{24 \pi}.
\end{align}
\noindent This is just a $c$-number, and in fact for linear $\omega$ this will be zero. Similarly the $[O_L,[O_R, O_L]]$ computation can be done and the boundary term will at most give a $c$-number.

\subsection{Proportionality to $[O_R, O_L]$}

Now, let us ask, when is $[O_R, [O_R, O_L]]$ proportional to $[O_R, O_L]$ (apart from additive $c$-numbers). Demanding this will ensure that Zassenhaus expansion of $e^{i \mathcal{O}t}$ in terms of $O_{L}, O_R$ becomes very simple.  This will give us a differential equation:
\begin{align}
\label{-24}
\omega= \alpha W(\omega, R) \implies f R' - R f'= \alpha \left( \omega R' - \omega' R\right) .
\end{align}
\noindent for proportionality constant $\alpha$. Similarly in the $[O_L, [O_R, O_L]]$ computation,
\begin{align}
\label{-26}
 i^2 \int_I dy \cdot [\omega(y) L'(y)-\omega'(y)L(y)]T(y)
\end{align}
\noindent is the non-trivial part, which we want to be proportional to $[O_R,O_L]$. Therefore, we demand
\begin{align}
    \omega = \beta W(\omega,L) =  \beta(\omega L'-\omega' L) \label{omega-l}
\end{align}
\noindent for another proportionality constant $\beta$. Using the $R(x), L(x)$ obtained by solving these equations together will ensure that all nested commutators of $O_{L}, O_R$ are proportional to $[O_R, O_L]$ for the case of $\overline{f} = \overline{L}=\overline{R}= 0$, and will factorize the expansion into just three terms: $e^{i O_{L} t}, e^{i O_{R} t}$, and $e^{\Omega(t) [O_R, O_L]}$ for some function $\Omega(t)$. Further, the central charge terms will all collect into a $c$-number of the form $e^{\lambda (t)}$ for some function $\lambda(t)$.

If we had chosen $f = L = R=0$ instead of $\overline{f} = \overline{R} =\overline{L}= 0$, it is easy to see the change would only be the commutator relation of stress tensor components; compared to \eqref{TT}, there would be an overall minus sign. Since $[T(x), \overline{T}(y)]=0$,  we can say with both $f$ and $\overline{f}$ non-zero that if $\overline{\omega} (x) = \overline{L}(y) \overline{R}'(y) - \overline{L}'(y) \overline{R}(y)$,
\begin{align}
\label{-33}
 \overline{\omega} = - \alpha W(\overline{\omega}, \overline{R}) = -\beta W(\overline{\omega}, \overline{L})
\end{align}
\noindent are the additional equations to solve to obtain $\overline{R}, \overline{L}$ to ensure all nested commutators are proportional to $[O_{R}, O_L]$ where $\alpha, \beta$ are the same as above.

\subsubsection*{Constant function $f$}

By using $\omega = LR' - RL' = f R' - R f'$ these become very complicated differential equations, e.g. $\omega = \alpha W(\omega, R)$ becomes
\begin{align}
\label{-34}
\alpha R(x)^2 f''(x)+R'(x) \left[ -\alpha R(x) f'(x)-f(x)\right ]-\alpha f(x) R(x) R''(x)+\alpha f(x)   R'(x)^2+R(x) f'(x)=0
\end{align}
\noindent which we will not discuss for general functions $f$ in this work. However for a constant $f$, we can get the solutions easily. Using \eqref{-24} and \eqref{omega-l}, we can write
\begin{align}
\label{calc-}
&\omega=\beta W(\omega, L) = \beta W(\omega, f) - \frac{\beta}{\alpha} \omega
\intertext{Thus,}
&   f' - \frac{\omega'}{\omega} f = \left(\frac{1}{\beta}+\frac{1}{\alpha}\right)
\intertext{For a constant $f$,  we know $\omega = fR'$, giving us}
& \frac{R'(x)}{R(x)}  = \frac{d}{dx}\log R(x) =  - \frac{1}{f} \left(\frac{\beta + \alpha}{\alpha \beta}\right)  \implies R(x) = A \exp\left( -\frac{x}{f} \left( \frac{\alpha+\beta}{\alpha \beta}\right) \right) \label{exp-sol}
\end{align}
\noindent but this is not the only solution. Using \eqref{-24} directly we can see that
\begin{align}
\label{anoth}
R(x) = \frac{x}{\alpha} + C
\end{align}
\noindent is another solution. In fact, if one requires that $[O_R + O_L, [O_R,O_L]]=0, \alpha = -\beta$, so the solution \eqref{exp-sol}  becomes trivial, leaving \eqref{anoth}. The condition $[O_R+O_L, [O_R,O_L]]=0$  holds for half-sided translations.

\subsection{Special case of half-sided translations}

\noindent Let us now set $f(x) = 2a, R(x) = x, a_L = 0$, and $a_R= 2a$ in \eqref{-8} for $b_L ,b_R \to \infty$. Comparing with \eqref{Gfinal} and \eqref{G'final}, it makes $O_R =G$ and $O_L=G'$, and gives
\begin{align}
\label{-28}
& \omega(x) = f R'(x) = 2a
\end{align}
\noindent Recall $\omega(x) = f R'(x)$ for constant $f$, making $\omega(x) = 2a$. Further, note that $R(x)$ is linear and $\omega$ is a constant, therefore there are no boundary terms, nor any terms involving central charge $c$.  Further, we can compute $W(\omega,L)= W(2a,2a-x)= -2a$, which is $-\omega(x)$ meaning that $\beta = -\alpha$, so that \eqref{exp-sol} is not a useful solution.

Taking care of the $i$ factors, it is straightforward to write all the nested commutators:
\begin{align}
\label{allne}
\text{ad}_G^{k} G' = i^{k-1}[G,G'], \quad \text{ad}_{G'}^{k} G' &= (-i)^{k-1} [G,G'], \quad \text{ad}_{\mathcal{G}}^{j} G' =  \delta_{j,1}\cdot[G,G'] ,\; j \geq 1
\intertext{ with}
 [G,G'] = i \int_0^{2a}& dx\cdot\omega(x) T(x) = 2a i\int_{0}^{2a} dx \cdot T(x). \label{old-prob}
\end{align}

\subsubsection{Regularization}
\label{ssec:regul}

Following our discussions in section \ref{sec:g-g'},  we see that equation \eqref{old-prob} clearly has problems coming from integral's boundaries $x=0$ and $x=2a$. Using the same logic as before, we can regularize it at the boundaries with functions $\mu(x) =  (x /\epsilon)^{1+\epsilon}$ on $(0,\epsilon)$ and $\nu(x) =2a - [ (2a-x)/\epsilon ]^{1+\epsilon}$ on $(2a-\epsilon, 2a)$ such that UV divergence vanishes, where this time $\epsilon \to 0$ makes $\mu, \nu$ diverge. However there is a problem with such a regularization; fixing $[G,G']$ does not fix any of the nested commutators such as $[G,[G,G']]$.

This can be seen from the general results of this section. For any  $R(x)$, we can compute $\omega(x) = f R'(x)$, and each time we nest commutators, we will have to differentiate $R(x)$. For any analytic function used for regularizing, there will be an $N$ such that $R^{ (N+1) }(\epsilon)$ is discontinuous. This follows from a simple Taylor expansion; demanding all derivatives of $R^{(n)}(x)$ vanish at $x=\epsilon$ for $n \geq 2$ will force $R(x)$ to be linear. This non-differentiability of $R(x)$ gives rise to singularities, because second derivative of a non-differentiable function has an infinite discontinuity. Thus, we see, that for defining commutators, we must ensure $R(x)$ post regulation is smooth, forcing us to go beyond analytic functions.

In the subject of smooth manifolds \cite{Lee:2012sm}, such functions do exist, and they often rely on the function $\exp( -1 /x )$, which has an essential singularity at $x=0$ when $x$ is treated as a complex variable.  For our purposes,\footnote{It is worth mentioning that there is a well known and simpler way to construct a smooth ramp function that tapers off to constant value using $f(x) = \{e^{-1 /x}, x\geq 0;\; 0, x\leq  0\}$ by considering a ratio like $f(2-x)/[f(2-x)+f(x-1)]$, but the this ramp does not have a uniform slope, and therefore does not work for our problem. } we can define the function
\begin{align}
\label{mumol}
\psi(t) =\frac{1}{C} \int_0^{t} ds\, \exp\left( -\frac{1}{s(1-s)} \right) , \quad C  = \int_0^{1} ds\, \exp\left( -\frac{1}{s(1-s)} \right) .
\end{align}
\noindent If we consider now a piecewise function
\begin{align}
\label{cases}
P(x) =
\begin{cases} 
    0, & x\leq  0 \\
   x \psi(\frac{x}{\epsilon}), & 0 \leq x\leq  \epsilon \\
\; x, & x\geq \epsilon
\end{cases}
\end{align}
\noindent it follows that $P^{(n)}(\epsilon) = 0$ for all $n \geq 2$, and $P^{ (n) }(0)=0$ for all $n\geq 0$. We can show this by noting that
\begin{align}
\label{deriv}
\psi'(t) = \frac{1}{C} \exp\left( - \frac{1}{t (1-t)} \right)  .
\end{align}
\noindent Thus instead of taking the minimal regularization of \eqref{minim}, we will instead require
\begin{align}
\label{regul}
\mu(x) = x \psi \left(\frac{x}{\epsilon}\right), \quad \nu(x) = 2a - (2a-x) \psi \left( \frac{2a-x}{\epsilon}\right)
\end{align}
\noindent in the expressions \eqref{hatg}. This modification for $G,G'$ will regularize all the nested commutators in one go. However, note that any regularization, including this one, will break down the proportionality of the nested commutators \eqref{allne}. This will prevent a simple factorization of $e^{i s \mathcal{G}}$ near the boundaries, but that will not be a major problem because we will be able to use \eqref{allne} away from the boundaries.

\section{Zassenhaus expansion for half-sided translations}
\label{sec:zas}

\noindent The Zassenhaus expansion \cite{Magnus:1954ex} is ubiquitous in quantum mechanics for writing operators $e^{\lambda (X+Y)}$ in terms of exponentials of $X$,  $Y$ and their commutators in a right sided expansion:
\begin{align}
\label{rse}
e^{\lambda (X+Y)} = e^{\lambda X} e^{\lambda Y} \cdot \prod_{i=2}^{\infty} e^{\lambda^{n} C_n}
\end{align}
\noindent for some $C_n$ which can be obtained recursively by differentiating the expression repeatedly. In general this is a difficult task, however, there exists an efficient recursive method for their computation due to Casas et al \cite{Casas:2012ecz}. Their result is
\begin{align}
&    C_{n} =
\begin{cases}
\frac{1}{n} f_{1, n-1},\quad\quad\quad n=2,3,4\\
\frac{1}{n} f\left({\left[\frac{n-1}{2}\right], n-1}\right) , \quad n \geq 5
\end{cases} \label{casas}
\intertext{where}
& f_{1,k} =   \sum_{j=1}^{k} \frac{(-1)^{k}}{j! (k-j)!} \text{ad}_{Y}^{k-j} \text{ad}_{X}^{j} Y,
\\ & f_{n,k} = \sum_{j=0}^{[k/ n] -1}  \frac{(-1)^{j}}{j!} \text{ad}^{j}_{C_n} f_{n-1, k-nj}, \quad k\geq n.
\end{align}
\noindent where $\text{ad}_X Y = [X,Y], \text{ad}_X^k Y = [X, \text{ad}^{k-1} _X Y]$, and $\text{ad}_X^0 Y = Y$.

\subsection{Issue with the right sided expansion}

We can demand a right sided expansion for half-sided translations, with $X = \hat{G}', Y = \hat{G}$ and $\lambda = is$,
\begin{align}
\label{rsere-2}
e^{i s \mathcal{G}} = e^{is \hat{G}'} e^{is \hat{G}} \prod_{n=2} ^{\infty} e^{(is)^{n} C_n}
\end{align}
\noindent where $C_n$ can be obtained using \eqref{casas} and turn out to be integrals of stress tensor over $(0,2a)$ that are closely related to \eqref{old-prob}. Can this expression describe half-sided translations? Let us ask this question by considering the conjugation for $\phi(x)$, for $x$ spacelike to the interval $(0,2a)$.
\begin{align}
\label{check}
A(s,x)= \exp( i s \mathcal{G} ) \phi(x) \exp( -i s \mathcal{G} )
\end{align}
\noindent Since $e^{i s \mathcal{G}}$ is unitary, and $x$ is spacelike to $(0,2a)$, by micro causality $\phi(x)$ commutes with all $C_n$, giving,
\begin{align}
\label{-43}
A(s,x) = e^{is \hat{G}'} e^{is \hat{G}}  \cdot  \phi(x) \cdot e^{-is \hat{G}} e^{-is \hat{G}'}
\end{align}
\noindent but from the discussions of section \ref{sec:conjugations} conjugation by $e^{is \hat{G}'}$ and $e^{is \hat{G}}$ can transport $\phi(x)$ to regions causally connected with $(0,2a)$. Without the contributions from $C_n$, $\phi(x)$ cannot cross the region 2 of figure \ref{fig:rindl} in a way $e^{i \mathcal{G} s} \phi(x) e^{-i \mathcal{G} s}$ does. Therefore, \eqref{rsere-2} cannot represent half-sided translations when excitations outside region 2 are involved.

\subsection{Derivation of centered Zassenhaus formula}

The key to avoid this problem is to transport $\phi(x)$ into the overlap region before the commutator terms act on it. This can be done using a centered expansion which will ensure that commutator terms always make their contribution. We therefore should look for the decomposition
\begin{align}
\label{cente}
e^{\lambda (X+Y)} = e^{\lambda X} \cdot  \prod_{n=2}^{\infty} e^{\lambda^{n} D_n}  \cdot e^{\lambda Y}.
\end{align}
Unfortunately the centered version of Zassenhaus expansion does not seem to be available in the literature and therefore we will have to derive it. We can take inspiration from \cite{Casas:2012ecz}, of proposing an expansion and taking derivatives to get $\{D_n\}$. We demand $e^{\lambda (X+Y)}$ has an expansion \eqref{cente}. Then let us define the functions
\begin{align}
\label{fulld}
&\quad R_1 (\lambda) = e^{-\lambda X}e^{\lambda(X+Y)} e^{-\lambda Y}, \quad F_1 = \frac{d}{d\lambda} R_1 \cdot  R_1^{-1}.
\intertext{Note that $R_1(\lambda)$ is equal to the infinite product $\prod_{n=2} ^{\infty} e^{\lambda^{n} D_n}$.
Using this, one can define}
&R_n (\lambda) = e^{-\lambda^{n}D_n} R_{n-1} (\lambda), \quad F_n = \frac{d}{d\lambda} R_n \cdot R_n ^{-1}.
\intertext{These functions like in \cite{Casas:2012ecz} obey the recursive relation}
&F_n(\lambda) = e^{-\lambda^{n} \text{ad}_{D_n} } (F_{n-1} (\lambda) - n D_n \lambda^{n-1}) \label{recur}
\intertext{which enable us to find $D_n$' s recursively, starting from $F_1$. Therefore, let us begin the task.}
&\frac{\partial R_1}{\partial \lambda} = -X \cdot R_1 - R_1 \cdot Y + e^{-\lambda X} (X+Y) e^{\lambda(X+Y)} e^{-\lambda Y}
\intertext{and now, we use}
&R_1'(\lambda) \cdot R_1^{-1} = -X -R_1 Y R_1^{-1} + e^{-\lambda X} (X+Y) e^{\lambda X}
\\ &= -X - e^{-\lambda X}e^{\lambda(X+Y)} Y e^{-\lambda(X+Y)}e^{\lambda X}+ e^{-\lambda X} (X+Y) e^{\lambda X}
\\&= -X +  e^{-\lambda X} \; [-e^{\lambda(X+Y)} Y e^{-\lambda(X+Y)} + X+Y] \; e^{\lambda X}
\\&=   e^{-\lambda X} \; [-e^{\lambda(X+Y)} Y e^{-\lambda(X+Y)}  +Y] \; e^{\lambda X}
\\&=   -e^{-\lambda \text{ad}_X}(e^{\lambda \text{ad}_{X+Y}} -I)Y
\\&=   -\sum_{j=0}^{\infty} \sum_{p=1}^{\infty} \frac{(-\lambda)^{j} \lambda^{p}}{j!p!} \text{ad}_{X}^{j} \text{ad}_{X+Y}^{p} Y
\intertext{ Like in \cite{Casas:2012ecz}, let us define}
& F_n (\lambda) = \sum_{k=n}^{\infty} f_{n,k} \lambda^{k}.
\intertext{From this, we can extract $f_{1,k}$ by a change of index; $\lambda^{j+p}$ must be set to $\lambda^{k}$,  and the sum is over $k=1$ to infinity.  Thus}
&f_{1,k} =  \sum_{j=0}^{k-1} \frac{(-1)^{j+1}}{j! (k-j)!} \text{ad}_{X}^{j} \text{ad}_{X+Y}^{k-j} Y.
\intertext{Using the recursive relation \eqref{recur}}
& f_{n,k} = \sum_{j=0}^{[k/n]-1} \frac{(-1)^{j+1}}{j!} \text{ad}_{D_n}^{j} f_{n-1, k-nj}
\intertext{One now finally obtains}
&D_{n} =
\begin{cases}
\frac{1}{n} f_{1, n-1},\quad\quad\quad n=2,3,4\\
\frac{1}{n} f\left({\left[\frac{n-1}{2}\right], n-1}\right) , \quad n \geq 5
\end{cases} \label{casas}.
\end{align}
\noindent Compared to the right sided expansion in \cite{Casas:2012ecz}, only thing that has changed is the form of $f_{1,k}$.

\subsubsection{Centered Zassenhaus expansion}

First let us do the calculation for the forms $G,G'$, then we will look at operators $\hat{G},\hat{G}'$. we will set $Y = G, X=G'$, and $\lambda = is$.
\begin{align}
\label{-6}
f_{1,k} &= \sum_{j=0}^{k-1} \frac{(-1)^{j+1}}{j! (k-j)!} \text{ad}_{G'}^{j} \text{ad}_{\mathcal{G}}^{k-j} G
\\  &=   \frac{(-1)^{k}}{(k-1)!} \text{ad}_{G'}^{k-1} \cdot [G',G]  = \frac{(-i)^{k-1} (-1)^{k-1} }{(k-1)!} [G,G'].
\end{align}
\noindent It is easy to see that $f_{n,k}=f_{1,k}$ because of this simplified structure, and in fact for all $n \geq 2$:
\begin{align}
\label{-7}
&D_n = \frac{1}{n} f_{1,n-1}  =   [G,G'] \cdot  \frac{i^{n-2}}{n (n-2)!}
\intertext{Therefore}
& \sum_{n=2}^{\infty} \lambda^{n} D_n = \frac{1}{i}[G,G'] \cdot \sum_{n=2}^{\infty}\frac{\lambda^{n} \cdot i^{n-1}}{n(n-2)!} = \frac{1}{i}[ e^{i\lambda} (\lambda +i)-i] [G,G']
\\ & =  [e^{-s} (s+1)-1] \cdot [G,G'].
\intertext{Therefore, we have the result}
\exp(is \mathcal{G}) &= e^{i s G'} \cdot  \exp( [e^{-s}(s+1)-1]\cdot  [G,G'] ) \cdot  e^{i s G} \label{hst-result}.
\end{align}

As a comment, this closed form answer resulted from all nested commutators being proportional to the most basic one. Such a closed form solution for a Zassenhaus expansion was shown to hold for a class of bounded operators $X, Y$ whose commutator was $[X,Y] = u X  + v Y + c 1$ in \cite{Dupays:2021cfz}. For half-sided translations, the bilinear forms $G,G'$ show some similarity to this prescription if we set $Y=G, X=G'$ with  $u =v=1$. That would mean $[G,G'] = \mathcal{G} + 1 c$. This is not true in our case for any $c$, but it is true that $[\mathcal{G}, [G,G']]=0$.

\subsubsection*{Upgrading to operators}

Unfortunately, for $\hat{G}, \hat{G}'$, such a simple form will not appear. One can however show that because of smoothness of bump regulators, the changes will only be near the boundary, giving
\begin{align}
\label{centr-2}
\exp(i s \mathcal{G}) =  e^{i s \hat{G}'} & \prod_{n=2}^{\infty}  \exp\left[ (is)^{n} \left(   \frac{i^{n-2}}{n(n-2)!} D(\epsilon)   +\Sigma_n(\epsilon)\right )\right ] \cdot e^{i s \hat{G}}
\\  &D(\epsilon) = 2ai \int_\epsilon^{2a-\epsilon} T(x) \, dx 
\end{align}
\noindent where the boundary terms $\Sigma_n(\epsilon)$ involve nested Wronskians of $\mu(x), \nu(x)$ with $R(x)$, and do not commute with the integral on $(\epsilon,2a-\epsilon)$. To see this, note that $\eqref{-6}$ now has contributions from  $k-j>1$, all of which give operators that commute with integral on  $(\epsilon, 2a-\epsilon)$. However, the $k-j=1$ contribution itself is such that the its boundary terms do not commute with the part on $(\epsilon, 2a-\epsilon)$, which is $D(\epsilon)$ for any $j$. 

Note that the boundary modifications can produce central charge factors, which can be estimated by numerical integrations of functions involving $\psi(t)$ in \eqref{mumol} over the intervals $(0,\epsilon)$ and $(2a-\epsilon,2a)$. However, because of the $e^{- 1/t}$ decay of $\psi(t)$ near $t=0$, they can be assumed to be negligible. Further, the explicit form of $\Sigma_{n}(\epsilon)$ can be worked out by taking Wronskians repeatedly but this task is neither straightforward nor very illuminating for our present study. 

The above result means that as long as $x$ remains is in $(\epsilon,2a-\epsilon)$, \eqref{centr-2} takes the form of \eqref{hst-result} for conjugation $e^{i s \mathcal{G}} \phi(x) e^{-i s \mathcal{G}}$. As a remark, in the above discussion we assumed $s$ was positive. However $s$ can be negative, and that case requires an interchange of $\hat{G},\hat{G}'$, or alternatively, taking the adjoint of the equation \eqref{centr-2}.

\subsection{Testing the result}

We will now verify that \eqref{centr-2} reproduces the effect of half-sided translations for fields that stay confined in region 2 and spacelike from $x=\epsilon, 2a-\epsilon$. In this case, we can ignore the $\Sigma_n(\epsilon)$ in each conjugation, and the product \eqref{centr-2} is effectively given by
\begin{equation}
    \exp(i s \mathcal{G}) \stackrel{*}{=} e^{i s \hat{G}'} \cdot \exp ( g(s) D(\epsilon)) \cdot e^{i s \hat{G}} 
\end{equation}
where $g(s)$ is the function from \eqref{hst-result}, and the $*$ is to denote that this holds only in our chosen case. Let's ask what effect  $e^{g(s)D(\epsilon)}$ has on a $\phi(w, \overline{w})$, a primary. This will be similar to our calculations in section \ref{sec:conjugations}.  We will need the $[T(z),\phi(x)]$ commutator, and first we are looking for 
\begin{align}
\label{middl}
& e^{g(s) D(\epsilon)} \phi (w,\overline{w}) e^{-g(s) D(\epsilon)}
\intertext{when $w$ is in $(\epsilon, 2a-\epsilon)$.  Using Hadamard lemma, this is same as }
& \sum_{n=1}^{\infty} \frac{g^{n}}{n!} \text{ad}_{D(\epsilon)}^{n} \phi.
\intertext{As we are away from the boundary, the boundary terms drop.}
& [D(\epsilon),\phi]=-2a i  (-i) \int dx \cdot \partial_x \phi(x,\overline{w}) \delta(x-w) =-2a  \cdot \partial_w\phi(w,\overline{w}).
\intertext{Thus,}
& \sum_{n=0}^{\infty} \frac{g^{n}}{n!} (-2a)^{n}\partial_w^{n} \phi(w,\overline{w}) = \phi(w-2a g(s), \overline{w}). \label{g-eff}
\end{align}
\noindent Thus, if initial location was $w=w_0$, after the conjugation by $e^{g(s)D(\epsilon)}$ the location is $w=w_0 - 2a g(s)$. We can now ask, what if we were to apply $e^{i\hat{G}'s}$ on this $\phi$. From \eqref{leftt}, we already know the final answer would be
\begin{align}
\label{verif}
\phi(w(s), \overline{w})& = \exp[sh+ s(w -2a)\partial_w] \cdot \phi(w,\overline{w})
\intertext{where the derivative is to be evaluated at $w_0 - 2ag(s)$. Like we did in section \ref{sec:conjugations}, using the change of variable $u = \ln (w-2a)$, this simplifies to a translation operation, giving}
\phi(w(s), \overline{w}) &= e^{sh} \phi(2a -e^{s}[ 2a - (w_0 - 2ag(s))], \overline{w}).
\intertext{This is the result of composing $e^{g(s)D(\epsilon)}$ and $e^{i \hat{G}'s}$ in correct sequence. To include the effect of $e^{i\hat{G}s}$ at the beginning, following \eqref{scale-dim}, we can do a rescaling $w_0 \to  e^{-s} w_0$  and cancel $e^{-sh}$. Composing all three therefore gives}
\phi(w(s),\overline{w}) = & e^{-sh + sh} \phi(w_0 + 2a (1-e^{s}) - 2ag(s) e^{s} ).
\intertext{Now finally if we set $g(s)= e^{-s}(s+1)-1$,  the answer is }
\phi(w(s),\overline{w}) & =  \phi(w_0 - 2as, \overline{w})
\end{align}
\noindent which is exactly what we will get from $e^{i \mathcal{G} s}$, because $w = x-t$ and it decreases as $t$ increases.

As a final remark, let us note that we were able to cancel the $e^{-sh}, e^{sh}$ factors because we confined $\phi(w(s),\overline{w})$ for all $s$ entirely to the region 2. This would not have happened otherwise, and this is one of the reasons why on its own, \eqref{hst-result} is not correct, and the boundary terms in \eqref{centr-2} are important. The figures \ref{fig:plots1} and \ref{fig:plots2} are polynomial approximations of the derivatives of $R(x)$ for half-sided translations with and without regularization with $\mu(x)$ and $\nu(x)$ of \eqref{regul}, which show the non-trivial edge effects that arise near the boundary.

\begin{figure}[ht]
\centering

\begin{tikzpicture}
  \begin{axis}[
    width=10cm,
    height=7cm,
    axis lines=left,
    xlabel={$x$},
    ylabel={$R'(x)$},
    ymin=0, ymax=1.0,
    xmin=-1.1, xmax=1.1,
    xtick={-1, -0.8,0,0.8,1.0},
    ytick={0,1},
    grid=none,
    ticklabel style={font=\small},
    label style={font=\small},
    legend pos=north west
  ]

  % Reference box function (black step)
  % \addplot[black, thick, domain=0:0.1] {10*x};
  \addplot[black, thick, domain=-1:1] {1/2};
  % \addplot[black, thick, domain=0.9:1] {10*(1 - x)};

  \addplot[dashed, gray] coordinates {(-0.8, -1) (-0.8, 1)};
  \addplot[thick, black] coordinates {(-1, -1) (-1, 1/2)};
  \addplot[thick, black] coordinates {(1, -1) (1, 1/2)};
  \addplot[dashed, gray] coordinates {(0.8, -1) (0.8, 1)};
  % Smooth derivative function R'(x)
  \addplot[red, thick, domain=-1:1, samples=500]{
    x < -0.8 ? 
      4687.5*x^4 + 16750*x^3 + 22350*x^2 + 13200*x + 2912.5 :
    (x > 0.8 ? 
     4687.5*x^4 - 16750*x^3 + 22350*x^2 - 13200*x + 2912.5  :
     1/2)
  };
  \legend{$R'(x)$}
  \end{axis}
\end{tikzpicture}
\caption{First derivative of $R(x)$ as a smoothed function with a 5th-degree polynomial on $(-1,1)$ with $\epsilon = 0.2$ shown in red; the original $R(x)\propto x$ having a constant derivative is shown in black.}
\label{fig:plots1}
\end{figure}

\begin{figure}[ht]
\centering

\begin{tikzpicture}
  \begin{axis}[
    width=10cm,
    height=7cm,
    xlabel={$x$},
    ylabel={$R''(x)$},
    xmin=-1.0, xmax=1.0,
    grid=both,
    grid style={dotted,gray!30},
    axis lines=left,
    enlargelimits=0.05,
    legend pos=south east,
    ticklabel style={font=\small},
    label style={font=\small},
    title={}
  ]

  \addplot[black, thick, domain=-1:1] {0};
  \addplot[dashed, gray] coordinates {(-0.8, -10) (-0.8, 10)};
  \addplot[dashed, gray] coordinates {(0.8, -10) (0.8, 10)};

  \addplot[blue, thick, domain=-1:1, samples=400]{
    x < -0.8 ? 
       18750*x^3 + 50250*x^2 + 44700*x + 13200 :
    (x > 0.8 ? 
       -18750*x^3 + 50250*x^2 - 44700*x + 13200 :
     0)
  };
  \addlegendentry{$R''(x)$}

  \end{axis}
\end{tikzpicture}
\caption{Second derivative of $R(x)$ as a smoothed function with a 5th-degree polynomial on $(-1,1)$ with $\epsilon = 0.2$.}
\label{fig:plots2}
\end{figure}

\section{Conclusions}

In this work we have studied Zassenhaus decomposition of exponential operators for 1+1 dimensional conformal field theories, with main emphasis on half-sided translations. We focused on the observation that one can write $\mathcal{G} = G+ G'$, where the right side is built out of entanglement Hamiltonians, however, $G,G'$ are not well-defined operators. We have shown how $G,G'$ can be replaced by operators $\hat{G}, \hat{G}'$ with well-defined nested commutators, and using a central Zassenhaus expansion, we have obtained a decomposition of $e^{i s \mathcal{G}}$ in terms of $\hat{G}, \hat{G}'$, and their commutators.

One speculative comment that may be relevant here is about $\hat{G}, \hat{G}'$ not being self-adjoint. In section \ref{sec:conjugations} we saw that conjugation of a primary $\phi$ by these operators does not preserve the operator norm. This is similar to what happens in quantum mechanics for a particle living on half space $(0, \infty)$. The operator $-i \frac{d}{dx}$ is not self adjoint on $\mathcal{L}^2(0, \infty)$, and therefore $e^{ \frac{d}{dx}}$, the translation operator, is not unitary. This can be seen through the fact that $e^{\frac{d}{dx}}$ maps the elements of $\mathcal{L}^2(0, \infty)$ to a proper subset by shifting them to right (the map is not onto). The map is in fact an isometry, and while these preserve the inner product, they are not expected to preserve operator norms on conjugation. The situation here is very similar, and it seems to suggest that can $e^{i \hat{G}s}, e^{i \hat{G}'s}$ may be isometries.

As a future direction it would be interesting to explore whether the class of operators permitting decomposition \eqref{-8} can be generalized to include light-ray operators \cite{Besken:2021oli}.  One interesting setup to study the decomposition in detail would be for the entanglement Hamiltonian corresponding to a causal diamond, which can be dealt with using the tools developed in this work. This one may be particularly interesting because of its possible connections with entanglement between subregions $(-b_L, a_R)$ and $(-a_L, b_R)$ in the limit of shrinking overlap region $(-a_L, a_R)$ in figure \ref{fig:these}. It would also be interesting to see whether the results can be extended to conformal field theory on AdS$_2$ for applications to JT gravity, and whether they have deeper connections going back to the split property \cite{Buchholz:1986ntq}.

%Discuss split property.. the functions's derivatives... how the boundary terms matter... in preventing a "return" to left edge.

\noindent $ $

\noindent \textbf{Acknowledgements}

\noindent $ $

\noindent I wish to thank A. P. Balachandran, Budhaditya Bhattacharjee, Adarsh Sudhakar, and Nemani Suryanarayana for their comments and suggestions. Further, I would like to express my gratitude to Ronak Soni for valuable discussions and comments on the work, and to C. H. Namitha and S. Sundar for helping me with some mathematical aspects of the work. Finally, I wish to thank to Gr\'egoire Mathys and Edward Witten for important clarifications.

\noindent \appendix

\section{Difference of boost operators to obtain $\mathcal{G}$ for Rindler wedge}

Let us use the Poincar\'e algebra generators to obtain difference between boost operators corresponding to different stationary points. The algebra is given by
\begin{align}
\label{diffe}
&[K, P]= iH,\quad [K,H]=iP, \quad [H,P]=0.
\intertext{The boost operator $K$ preserves the origin $(t,x)=(0,0)$, thus the boost transformation that preserves $(t,x)=(v_t,v_x)$ is given by}
&e^{i t K(v_t, v_x)} = e^{i(v_t H - v_x P) }e^{i K t} e^{-i (v_t H - v_x P)}.
\intertext{One can expand $e^{iKt}$ in a Taylor series, and use Hadamard lemma and the Poincar\'e algebra repeatedly to get the answer.  The first commutator in the series is}
&[v_t H - v_x P, K]=-i(v_t P -v_x H).
\intertext{Using commutation relations,}
&[v_t H-v_x P, [v_t H-v_x P,
P. K]]=-i[v_t H -v_x P, v_t P -v_x
P. H]= 0.
\intertext{The nested commutators vanish, and we get}
&K(v_t, v_x)=K + (v_t P-v_x H).
\intertext{Because of the linearity of vector $v$ in this expression, we see for any $(x_0,t_0)$,  }
&\mathcal{G} = 2 a P^{+}\equiv a (H +P).
\end{align}
We can use another method of direct subtraction and using properties of stress tensor: 
\begin{align}
\label{-61}
&\mathcal{G}= \frac{1}{2 \pi} \left( K_{\mathcal{M}} - K_{\mathcal{N}}\right)  = \int (x-x_0) T_{00} (t_0,x) dx - \int(x-x_0-a) T_{00}(t_0-a, x)dx  \\
\intertext{We will restrict to conformal field theories, for which we can write $T_{00} (t,x)=T(x-t)+\overline{T}(x+t)$, that yields}
&\mathcal{G}^{-}=\int(x-x_0)[T(x-t_0)-T(x-t_0+2a)]dx
=2a \int T(x-t_0) dx
\intertext{where we changed the variable $x \to x-2a$ in the last step for the second integral. Using  the fact that our conventions lead to $T(x)=T_{--}(x)$,  this becomes}
&=2a\int_{-\infty}^{\infty} T_{--}(x) dx  \equiv 2a P^+.
\end{align}
\bibliography{references.bib}
\bibliographystyle{JHEP} 

\end{document}